\documentclass[a4paper,12pt]{article}
\usepackage{amsthm}
\usepackage{amsmath}
\usepackage{amssymb}
\usepackage{color}
\usepackage{graphicx}
\usepackage{subfigure}
\usepackage{cite}
\usepackage[linktocpage]{hyperref}
\usepackage{setspace}

\usepackage{amssymb}
\usepackage{graphicx}
\usepackage{epsfig}
\usepackage{bm}
\usepackage{mathrsfs}
\usepackage{amsfonts}
\usepackage{epstopdf}
\usepackage{bm}
\usepackage{amsthm}
\usepackage{multirow}
\usepackage{subfigure}
\usepackage{amsmath}
\usepackage{textcomp}
\usepackage{fancyhdr}
\usepackage{mathtools}

\begin{document}
\let\endtitlepage\relax

\begin{titlepage}
\begin{center}

\newcommand\blfootnote[1]{%
	\begingroup
	\renewcommand\thefootnote{}\footnote{#1}%
	\addtocounter{footnote}{-1}%
	\endgroup
}

\vspace*{-1.0cm}
\renewcommand{\baselinestretch}{1.0}  
\setstretch{1.6}
{\Large{\textbf{Conserved charges and asymptotic symmetries of BTZ-like black holes in Einstein-bumblebee gravity}}}

\renewcommand{\baselinestretch}{1.0}  
\setstretch{1.2}

\vspace{8mm}
\centerline{\large{Hai-Feng Ding\blfootnote{Email:~haifeng1116@qq.com} }}
\vspace{-1mm}
\normalsize
\textit{Institute of Geometric Arts, Chuxiong Normal University, Chuxiong, Yunnan 675000, China}

\renewcommand{\baselinestretch}{1.0}  
\setstretch{1}


\end{center}
\vspace*{0cm}
\end{titlepage}
\vspace*{-0mm}
\begin{abstract}
	Using the solution phase space method, we investigate the conserved charges of BTZ-like black holes in Einstein-bumblebee gravity. Our study shows that the black hole mass, angular momentum and entropy are influenced by the Lorentz-violating parameter. Through the study of the AdS/CFT correspondence, we derive the asymptotic charge algebra, which consists of two copies of the Virasoro algebra with non-trivial central charges. By employing the Cardy formula, we calculate the microscopic entropy of dual conformal field theory, which precisely matches with the Bekenstein-Hawking entropy. Furthermore, by imposing the Detournay-Smoes-Wutte boundary conditions in the near horizon geometry of the extremal BTZ-like black hole, we obtain a Virasoro-Kac-Moody $\mathrm{U(1)}$ algebra, representing the symmetry algebra of a warped conformal field theory.   
\end{abstract}

\newpage

\section{Introduction}
Symmetry is fundamental in modern physics. According to the well-known Noether’s theorem, symmetries are related to conservation laws \cite{Noether:1918zz}. In the realm of black hole physics, the analysis of the asymptotic symmetry of spacetime geometry under appropriate boundary conditions brings new comprehension and great importance. The investigation of asymptotic symmetries in gravitational theories started from the works of Bondi-Metzner-Sachs (BMS) in 1962 \cite{Bondi:1962px,Sachs:1962wk}, which identified the BMS group of supertranslations and Lorentz transformations as asymptotic symmetry group (ASG). Another landmark discovery by Brown and Henneaux \cite{Brown:1986nw} established a definitive link between the asymptotic symmetries of three-dimensional asymptotic anti-de-Sitter ($\mathrm{AdS_3}$) spacetime and the symmetries of two-dimensional conformal field theory (CFT), serving as an early precursor of the AdS/CFT correspondence \cite{Maldacena:1998bw}. By applying the AdS/CFT to the Banados-Teitelboim-Zanelli (BTZ) black holes \cite{Banados:1992wn,Banados:1992gq}, the microscopic origin of the Bekenstein-Hawking entropy was derived \cite{Strominger:1997eq}. A parallel generalization of the AdS/CFT correspondence in near horizon region is the Kerr/CFT correspondence \cite{Guica:2008mu}, which links near horizon geometry of a four-dimensional extremal Kerr black hole with a two-dimensional chiral CFT. Another holographic realization beyond AdS/CFT is the $\mathrm{WAdS_3/WCFT}$ correspondence, relating warped $\mathrm{AdS_3}$ ($\mathrm{WAdS_3}$) to warped CFT (WCFT) \cite{Detournay:2012pc}.

In the theories of gravity, conserved charges such as black hole mass, angular momentum, electric charge and black hole entropy associated with exact and asymptotic symmetries have been extensively studied by various methods (see \cite{Compere:2018aar,Adami:2017phg,Frodden:2019ylc} for recent reviews of the methods for conserved charges). Based on the Noether procedure, the covariant phase space method (CPSM) was proposed by Wald and Iyer (also called Wald’s formalism) \cite{Lee:1990nz,Wald:1993nt,Iyer:1994ys}. In the framework of CPSM, a systematic approach for calculating conserved charges in generally covariant theories of gravity was developed by K. Hajian \textit{et al}, which is called solution phase space method (SPSM) \cite{Hajian:2015xlp}. The SPSM is a restriction of CPSM to solution space manifold, focusing on parametric variation and exact symmetries. In SPSM, the charge variations are computed by an integration over an arbitrary smooth, closed, and compact codimension-2 surface surrounding the black hole singularity. For reviews and applications of SPSM see Refs.\cite{Hajian:2016kxx,Ghodrati:2016vvf,Ding:2020bwa,Guo:2024zaa}. The study of conserved charges and asymptotic symmetries is crucial for understanding the thermodynamic properties and holographic dualities  in gravitational theories. Generally, a physical system is governed by its action and the boundary conditions; the former determines the equations of motion (and therefore the dynamics of the system). This also implies that different actions and boundary conditions contain different conserved charges and ASG. 

As a special family of modified gravitational theories, the Lorentz-violating (LV) theories of gravity \cite{Mattingly:2005re}, including Ho\v rava-Lifshitz theory \cite{Horava:2009uw}, ghost condensation \cite{Arkani-Hamed:2003pdi}, warped brane world \cite{Frey:2003jq,Cline:2003xy}, Einstein-aether theory \cite{Jacobson:2000xp,Eling:2004dk,Jacobson:2007veq} and Einstein bumblebee gravity \cite{Kostelecky:1988zi,Kostelecky:1989jw,Kostelecky:2003fs,Kostelecky:2010ze}, have attracted increasing interests. In the action of the bumblebee gravity, the spontaneous Lorentz violation arises from a potential $V(B_{\mu} B^{\mu})$ acting on a bumblebee vector field $B_{\mu}$ \cite{Kostelecky:2003fs,Kostelecky:2010ze}. In 2017, Casana \textit{et al} obtained an exact Schwarzschild-like black hole solution in bumblebee gravity \cite{Casana:2017jkc}. Subsequently, a number of exact solutions are obtained, including Schwarzschild-anti-de-Sitter (AdS)-like black holes \cite{Maluf:2020kgf}, slowly rotating Kerr-like black holes \cite{Ding:2019mal}, rotating BTZ-like black holes \cite{Ding:2023niy}, and other solutions \cite{Santos:2014nxm,Jha:2020pvk,Filho:2022yrk,Xu:2022frb,Liu:2024axg} within bumblebee gravity. Comprehensive analyses of the properties of these black holes have conducted, including gravitational lensing \cite{Ovgun:2018ran},  quasinormal modes \cite{Oliveira:2021abg}, Hawking radiation \cite{Kanzi:2019gtu}, black hole shadows \cite{Ding:2019mal,Wang:2021irh,Islam:2024sph,Ali:2024ssf,Pantig:2025bpd}, accretion disk \cite{Liu:2019mls}, superradiant instability \cite{Jiang:2021whw}, parameters constraint \cite{Wang:2021gtd,Gu:2022grg}. All of these properties exhibit effects induced by LV. Therefore, whether the LV effects in the bumblebee gravity influence the conserved charges and ASG remains an open question worthy of further study.

Although the Lorentz symmetry is broken in the theories of bumblebee gravity, it remains generally invariant under diffeomorphism transformation \cite{Bluhm:2007bd}. Therefore, the SPSM is still accommodation. In this paper, we employ the SPSM to investigate the conserved charges of a rotating BTZ-like black hole in Einstein-bumblebee gravity \cite{Ding:2023niy}. We compute the black hole mass, angular momentum and entropy, showing that these quantities are affected by LV parameter. As a consistent check, we also derive the first law of black hole thermodynamics. Mostly, the ASG analysis is carried out in pure gravity \cite{Compere:2008cv,Compere:2009zj,Blagojevic:2009ek,Detournay:2023zni,Godet:2020xpk}. In this work, we investigate the ASG in the gravity coupled with the bumblebee field non-minimally. Given that the black hole solution describes an asymptotic AdS spacetime, we study the AdS/CFT correspondence and obtain the asymptotic charge algebra, which consists of two copies of the Virasoro algebra with non-trivial central charges. Using the Cardy formula \cite{Cardy:1986ie}, we compute the microscopic entropy of dual CFT and find that it is in precise agreement with the Bekenstein-Hawking entropy. It is of interest to expand the scope of holography beyond asymptotically $\mathrm{AdS}$ spacetime. The most notable non-$\mathrm{AdS}$ backgrounds include $\mathrm{WAdS_3}$ spacetimes and near horizon geometry of extremal Kerr black holes (NHEK), these geometries have $SL(2, \mathbb{R}) \times U(1)$ isometries \cite{Anninos:2008fx,Bardeen:1999px}. The boundary conditions can be chosen so that the symmetries are enhanced to a semi-direct sum of Virasoro algebra and Kac-Moody $\mathrm{U(1)}$ algebra, which is the symmetry algebra of WCFT \cite{Detournay:2012pc}, suggesting that WCFT could be a holographic counterpart to $\mathrm{WAdS_3}$ or NHEK \cite{Compere:2008cv,Compere:2009zj,Blagojevic:2009ek,Detournay:2023zni,Hofman:2014loa,Aggarwal:2019iay,Detournay:2019xgl}. Besides being characterized by Virasoro-Kac-Moody $\mathrm{U(1)}$ algebra, WCFT is a two-dimensional non-relativistic quantum field theory (QFT) that exhibits Lorentz symmetry breaking. Therefore, it is important to explore a potential dual holographic model which violates Lorentz symmetry on both the gravity and the field theory sides, as well as to investigate the influence of LV effects on the structure of ASG and the central charges. Recently, a new class of boundary conditions was devised by Detournay-Smoes-Wutte to study asymptotic symmetries in near horizon geometry \cite{Detournay:2023zni}, which is a higher dimensional uplift of Godet-Marteau boundary conditions in two-dimensional gravity \cite{Godet:2020xpk}. An application of these new boundary conditions is given in Ref.\cite{Jiang:2024tmh}. In near horizon region, we apply the Detournay-Smoes-Wutte boundary conditions to study the asymptotic symmetries for near horizon geometry of extremal BTZ-like black hole. As a result, we obtain a Virasoro-Kac-Moody $\mathrm{U(1)}$ algebra, this provides a first step towards building a holographic dual for near horizon geometry of bumblebee black holes.

This paper is organized as follows. In Sec.2, we provide a brief review of the SPSM. In Sec.3, we derive the Noether-Wald charge density and the surface charge density for Einstein-bumblebee gravity. In Sec.4, we calculate the conserved charges and check the first law of thermodynamics for rotating BTZ-like black hole. In Sec.5, we study the asymptotic symmetries in asymptotic $\mathrm{AdS_3}$ background. Two copies of Virasoro charge algebra will be obtained. Subsequently, we derive the microscopic entropy of dual CFT. In the near horizon geometry of extremal BTZ-like black hole, we apply the new boundary conditions to obtain the symmetry algebra of WCFT. Finally, Sec.6 concludes with a summary of our main findings and outlooks.

\section{Solution phase space method}
The SPSM \cite{Hajian:2015xlp} is based on the CPSM \cite{Lee:1990nz,Wald:1993nt,Iyer:1994ys}, so firstly we give a brief review of CPSM to achieve the conserved charges for generic gauge theories.

\textbf{Covariant phase space method:} A phase space is a manifold $\mathcal{M}$ equipped with a symplectic $2$-form $\Omega$. We consider an $n$-dimensional generally covariant theory of gravity described by the action
\begin{equation}
	S(\Phi)=\int \mathbf{L}(\Phi),
\end{equation} 
where $\mathbf{L}$ is the Lagrangian $n$-form, and $\Phi$ denotes collectively all the dynamical fields. The variation of $\mathbf{L}$ leads to
\begin{equation}\label{key}
	\delta \mathbf{L}(\Phi)=\mathbf{E}_{\Phi} \delta \Phi+\mathrm{d} \mathbf{\Theta}(\delta \Phi, \Phi) ,
\end{equation}
where $\delta \Phi$ is a generic field variation and provides the basis for tangent bundle of the phase space $\mathcal{M}$. The set $\mathbf{E}_{\Phi}=0$ gives the equations of motion (EOM). $\mathbf{\Theta}$ is an $(n-1;1)$-form called Lee-Wald symplectic potential, where the $(n-1)$-form is over the spacetime and the $1$-form is on the tangent bundle of $\mathcal{M}$. The Lee-Wald symplectic form \cite{Lee:1990nz} is defined as
\begin{equation}\label{key}
	\Omega \left(\delta_{1} \Phi, \delta_{2} \Phi, \Phi\right)=\int_{\Sigma} \bm{\omega} \left(\delta_{1} \Phi, \delta_{2} \Phi, \Phi\right) ,
\end{equation}
where
\begin{equation}\label{key}
	\bm{\omega} \left(\delta_{1} \Phi, \delta_{2} \Phi, \Phi \right) \equiv \delta_{1} \mathbf{\Theta} \left(\delta_{2} \Phi, \Phi\right)-\delta_{2} \mathbf{\Theta} \left(\delta_{1} \Phi, \Phi\right)
\end{equation}
representing the symplectic current form, which is an $(n-1;2)$-form. $\Sigma$ is a smooth codimension-$1$ surface.

When $\Phi$ solves the EOM $\mathbf{E}_{\Phi}=0$ and $\delta \Phi$ solves the linearized EOM $\delta \mathbf{E}_{\Phi}=0$,  i.e., on-shell, the conserved condition
\begin{equation}\label{key}
	\mathrm{d} \bm{\omega} \left(\delta_{1} \Phi, \delta_{2} \Phi, \Phi\right) \approx 0
\end{equation}
is satisfied, where $\approx$ denotes the on-shell equality.

When the $U(1)$ gauge field $A$ presents in $\mathbf{L}$, in addition to the diffeomorphism generated by a vector field $\xi$, the Lagrangian remains gauge invariant under the transformation $A \rightarrow A+\mathrm{d} \lambda$ for an arbitrary scalar $\lambda$. Consequently, the general gauge transformation takes the form $\delta_{\epsilon} \Phi=\left\{\mathcal{L}_{\xi} \Phi, \delta_{\lambda} \Phi\right\}$ where the generator $\epsilon$ is a combination of diffeomorphism and gauge transformation, given by $\epsilon \equiv\left\{\xi, \lambda\right\}$.

For field-dependent transformations, i.e. $\delta \epsilon \neq 0$, the charge variation $\delta H_{\epsilon}(\Phi)$ associated with generators $\epsilon$ can be defined as
\begin{align} \label{eqdh}
	\delta H_{\epsilon}(\Phi) & \equiv \Omega\left(\delta \Phi, \delta_{\epsilon} \Phi, \Phi\right)=\int_{\Sigma} \bm{\omega}\left(\delta \Phi, \delta_{\epsilon} \Phi, \Phi\right)  \nonumber \\  &\equiv \int_{\Sigma}\left(\delta^{[\Phi]} \mathbf{\Theta}\left(\delta_{\epsilon} \Phi, \Phi\right)-\delta_{\epsilon} \mathbf{\Theta}(\delta \Phi, \Phi)\right) \nonumber \\ &=\oint_{\partial \Sigma} \bm{k}_{\epsilon}(\delta \Phi, \Phi) ,
\end{align}
where
\begin{equation}\label{key}
	\bm{\omega}\left(\delta \Phi, \delta_{\epsilon} \Phi, \Phi\right) \approx \mathrm{d} \bm{k}_{\epsilon}(\delta \Phi, \Phi),
\end{equation}
and
\begin{equation}\label{k expression}
	\bm{k}_{\epsilon}(\delta \Phi, \Phi)=\delta \mathbf{Q}_{\epsilon}-\xi \cdot \mathbf{\Theta}(\delta \Phi, \Phi) .
\end{equation}
Here $\bm{k}_{\epsilon}$ is an $(n-2;1)$-form, called the surface charge density. The $(n-2;2)$-form $\mathbf{Q}_{\epsilon}$ is Noether-Wald charge density, related to the Noether current $\mathbf{J}_{\epsilon}$ by $\mathbf{J}_{\epsilon} \approx \mathrm{d} \mathbf{Q}_{\epsilon}$, and
\begin{equation}\label{NW charge density}
	\mathrm{d} \mathbf{Q}_{\epsilon} \equiv \mathbf{\Theta} \left(\delta_{\epsilon} \Phi, \Phi\right)-\xi \cdot \mathbf{L} .
\end{equation}
In Eq.(\ref{eqdh}), $\delta^{[\Phi]}$ denotes that $\delta$ acts only on $\Phi$ but not on the $\epsilon$ inside $\bm{\Theta}$.

If $\delta H_{\epsilon}(\Phi)$ is well-defined and integrable, we could determine the conserved charge $H_{\epsilon}$. The integrability condition is basically $\left(\delta_{1} \delta_{2}-\delta_{2} \delta_{1}\right) H_{\epsilon}(\Phi)=0$, which is equivalent to \cite{Hajian:2015xlp}
\begin{equation}\label{intcondition}
	\oint_{\partial \Sigma}\left(\xi \cdot \bm{\omega}\left(\delta_{1} \Phi, \delta_{2} \Phi, \Phi\right)+\bm{k}_{\delta_{1} \epsilon}\left(\delta_{2} \Phi, \Phi\right)-\bm{k}_{\delta_{2} \epsilon}\left(\delta_{1} \Phi, \Phi\right)\right) \approx 0 .
\end{equation}

To ensure the conservation of charge variation $\delta H_{\epsilon}(\Phi)$ and the independence of integration on $\Sigma$ and $\partial \Sigma$, the symplectic current form $\bm{\omega}$ must vanish on-shell for a certain subclass of $\delta_{\epsilon} \Phi$'s, i.e.,
\begin{equation}\label{symplectic symmetry}
	\bm{\omega}\left(\delta \Phi, \delta_{\epsilon} \Phi, \Phi\right) \approx 0 .
\end{equation}
In this case the transformations generated by $\delta_{\epsilon} \Phi$ are called symplectic symmetries. The family of $\epsilon$'s with this property can be divided into two sets: (1) non-exact symmetry generators denoted by $\chi$, for which $\delta_{\chi} \Phi \neq 0$ at least on one point in the phase space. (2) exact symmetry generators denoted by $\eta$, for which $\delta_{\eta} \Phi = 0$ all over the phase space.

\textbf{Solution phase space method:} The SPSM is a specification of CPSM to solution space manifold, which deals with parametric variation and exact symmetries $\hat{\delta}_{\eta} \Phi=0$. The solution space manifold $\hat {\mathcal{M}}$ is composed of solutions $\hat {\Phi}=\hat{\Phi}\left(x ; p_{\alpha}\right)$, where $p_{\alpha}$ is a set of solution parameters. The parametric variations $\hat{\delta} \Phi$ are defined as
\begin{equation}\label{key}
	\hat{\delta} \Phi \equiv \frac{\partial \hat{\Phi}}{\partial p_{\alpha}} \delta p_{\alpha} .
\end{equation}
The corresponding symplectic $2$-form $\Omega$ is denoted as $\hat\Omega$. The solution phase space is denoted by $( \hat{\mathcal{M}},\hat{\Omega})$, which is a submanifold of $( \mathcal{M},\Omega)$ when $\{\Phi \}$ are restricted to solution space $\{\hat\Phi \}$. The charge variations associated with the generators $\epsilon$ are taken as
\begin{equation}\label{dhintgration}
	\hat{\delta} H_{\epsilon}(\Phi)=\oint_{\partial \Sigma} \bm{k}_{\epsilon}(\hat{\delta} \Phi, \hat{\Phi}).
\end{equation}
The integrability condition (\ref{intcondition}) for parametric variations is given by
\begin{equation}\label{intcondition2}
	\oint_{\partial \Sigma}(\xi \cdot \bm{\omega}(\hat{\delta}_{1} \Phi, \hat{\delta}_{2} \Phi, \hat{\Phi})+\bm{k}_{\hat{\delta}_{1} \epsilon}(\hat{\delta}_{2} \Phi, \hat{\Phi})-\bm{k}_{\hat{\delta}_{2} \epsilon}(\hat{\delta}_{1} \Phi, \hat{\Phi})) \approx 0 .
\end{equation}

If the integration of Eq.(\ref{dhintgration}) is well-defined and integrable over parameters $p_{\alpha}$, the conserved charges $H_{\epsilon}\left(p_{\alpha}\right)$ could be calculated by
\begin{equation}\label{Hexpression}
	H_{\epsilon}( \hat{\Phi} (p_{\alpha}),\bar{\Phi} (\bar{p}_{\alpha}) )=\int_{\bar{p}}^{p} \hat{\delta} H_{\epsilon}+H_{\epsilon} ( \bar{\Phi} (\bar{p}_{\alpha}) ) ,
\end{equation}
where the integration is performed over arbitrary integral curves connecting a reference (or background) field configuration $\bar{\Phi} (x^{\mu}; \bar{p}_{\alpha})$ to the target space $\Phi (x^{\mu}; p_{\alpha})$. $H_{\epsilon}( \bar{\Phi} (\bar{p}_{\alpha}) )$ is the reference point for $H_{\epsilon}$ defined on $\bar{\Phi} (x^{\mu}; \bar{p}_{\alpha})$. 

\textbf{Algebra of surface charges:} When the charge variation $\delta H_{\epsilon}(\Phi)$ is well-defined and integrable, the algebra of the asymptotic symmetries is given by the Poisson bracket algebra of the surface charges \cite{Barnich:2007bf,Compere:2007az}
\begin{align}\label{poissonb}
	i\left\{H_{\epsilon} (\Phi,\bar{\Phi}),H_{\tilde{\epsilon}} ( \Phi,\bar{\Phi} ) \right\} &\equiv i \delta_{\tilde{\epsilon}} H_{\epsilon} (\Phi )  \nonumber \\
	&=i\oint_{\partial \Sigma}\bm{k}_{\epsilon}(\delta_{\tilde{\epsilon}} \Phi,\Phi)  \nonumber \\ &=i H_{ [ \epsilon,\tilde{\epsilon} ]} (\Phi,\bar{\Phi}) -i H_{ [ \epsilon,\tilde{\epsilon} ] } (\bar{\Phi}) +i \oint_{\partial \Sigma}\bm {k}_{\epsilon}(\delta_{\tilde{\epsilon}} \bar{\Phi},\bar{\Phi}),
\end{align}
where $[ \epsilon,\tilde{\epsilon} ]$ denotes the Lie bracket of the generator $\epsilon$, $i$ is the imaginary unit, and the last term $i \oint_{\partial \Sigma}\bm {k}_{\epsilon}(\delta_{\tilde{\epsilon}} \bar{\Phi},\bar{\Phi})$ is the central extention term in the algebra.
\section{Einstein-bumblebee theory}

In the bumblebee theory of gravity, bumblebee vector field $B_{\mu}$ is introduced to couples with gravity non-minimally. This field acquires a nonzero vacuum expectation value, leading to the spontaneous breaking of Lorentz symmetry in the gravitational sector. For Einstein-bumblebee gravity, the dynamical fields $\Phi$ consist of the metric $g_{\mu \nu}$ and bumblebee field $B_{\mu}$. The Lagrangian is given by \cite{Ding:2023niy}
\begin{equation}\label{Lagrangian}
	L=\frac{1}{2 \kappa}\left(R-2 \Lambda \right)+\frac{\rho}{2 \kappa} B^{\mu} B^{\nu} R_{\mu \nu}-\frac{1}{4} B_{\mu \nu} B^{\mu \nu}-V(B_{\mu} B^{\mu} \pm b^{2}) ,
\end{equation}
where $R$ is the Ricci scalar, and $\Lambda=-1/l^2$ is the negative cosmological constant. $\kappa=4 \pi G$ is the gravitational coupling constant, with $G$ being the Newtonian constant. $\rho$ is the nonminimal coupling constant between gravity and the bumlebee field. Analogous to the electromagnetic field, the bumblebee field strength is defined as
\begin{equation}\label{key}
	B_{\mu \nu}=\nabla_{\mu} B_{\nu}-\nabla_{\nu} B_{\mu} .
\end{equation}
The potential $V(B_{\mu} B^{\mu} \pm b^{2})$ triggers Lorentz violation, leading to a nonvanishing vacuum expectation value for $B_{\mu}$, such that $\left< B^{\mu} \right>=b^{\mu}$, where the vector $b_{\mu}$ satisfies $b^{\mu}b_{\mu}=\mp b^2=const$.

The Lagrangian $n$-form is the Hodge dual of $L$,
\begin{equation}\label{key}
	\mathbf{L}=\star{L}=L\, \bm{\epsilon} ,
\end{equation}
where $ \bm{\epsilon}=\sqrt{-g}\,d^nx=\frac{\sqrt{-g}}{n!} \varepsilon_{\mu_{1} \cdots\mu_{n}} d x^{\mu_{1}} \wedge \cdots \wedge d x^{\mu_{n}} $ is the $n$-dimensional volume form, $g$ is the determinant of the metric $g_{\mu \nu}$, and $\varepsilon_{\mu_{1} \cdots\mu_{n}}$ is the Levi-Civita symbol.
The variations of the Lagriangian $n$-form $\mathbf{L}$ with respect to $g_{\mu \nu}$ and $B_{\mu}$ are
\begin{equation}\label{key}
	\delta \mathbf{L}(\Phi)=\left( \mathrm{E}^{\mu \nu}_{g} \delta g_{\mu \nu}+\mathrm{E}^{\mu}_{B} \delta B_{\mu} \right)\bm{\epsilon} +\mathrm{d} \mathbf{\Theta}(\delta \Phi, \Phi) ,
\end{equation}
where the EOMs
\begin{align}
	\mathrm{E}^{\mu \nu}_{g}=&\frac{1}{2 \kappa} \left[-(G^{\mu \nu}+\Lambda g^{\mu \nu}) \right] \nonumber \\ +&\frac{\rho}{2 \kappa} \bigg[ \frac{1}{2} g^{\mu \nu} B^{\sigma} B^{\rho} R_{\sigma \rho} -2B^{(\nu} R^{\mu )}{}_{\sigma} B^{\sigma} + \nabla_{\sigma}\nabla^{(\mu}(B^{\nu )} B^{\sigma}) \nonumber \\ -&\frac{1}{2}\nabla_{\sigma}\nabla^{\sigma}(B^{\mu} B^{\nu})-\frac{1}{2}g^{\mu \nu} \nabla_{\sigma}\nabla_{\rho}(B^{\sigma}B^{\rho}) \bigg] \nonumber \\ +&\frac{1}{2} \left[ B^{(\mu}{}_{\sigma} B^{\nu ) \sigma}-\frac{1}{4} g^{\mu \nu} B^{\sigma \rho} B_{\sigma \rho}-g^{\mu \nu}V+2V'B^{\mu}B^{\nu} \right], \nonumber \\ \mathrm{E}^{\mu}_{B}=&\frac{\rho}{2 \kappa} \left[2 g^{\mu\nu} B^{\sigma} R_{\nu \sigma} \right]+\nabla_{\nu}B^{\nu \mu}-2 V'B^{\mu}, \label{Elambda}
\end{align}
in which
\begin{align}
	G_{\mu \nu}=&R_{\mu \nu}-\frac{1}{2} R g_{\mu \nu},  \nonumber \\
	V'=&\frac{\partial V(X)}{\partial X}, \nonumber \\
	X=&B_{\mu}B^{\mu} \pm b^2.
\end{align}

The symplectic potential $\mathbf{\Theta}(\delta \Phi, \Phi)$ is given by
\begin{equation}\label{key}
	\mathbf{\Theta}(\delta \Phi, \Phi)=\Theta^{\mu}(\delta \Phi, \Phi) \sqrt{-g}(d^{n-1}x)_{\mu} ,
\end{equation}
in which
\begin{equation}\label{key}
	\Theta^{\mu}(\delta \Phi, \Phi)=\frac{1}{2 \kappa} \left[ 2 \nabla^{[\nu } {h^{\mu ]}}_{\nu} \right]+\frac{\rho}{2 \kappa} \left[ \nabla_{\nu}(\Omega^{\mu \nu \sigma \rho} \delta g_{\sigma \rho})+\Sigma^{\mu \sigma \nu} \delta g_{\sigma \nu}\right]-B^{\mu \nu}\delta B_{\nu},
\end{equation}
with
\begin{align}
	\Omega^{\mu \nu \rho \sigma}=&g^{\mu (\rho}B^{\sigma)}B^{\nu}-\frac{1}{2}g^{\mu \nu}B^{\sigma}B^{\rho}-\frac{1}{2}g^{\sigma \rho}B^{\mu}B^{\nu}, \nonumber\\
	\Sigma^{\mu \sigma \nu}=&-g^{\mu (\nu}\nabla_{\rho}(B^{\sigma)}B^{\rho})-\nabla^{(\sigma}(B^{\nu)}B^{\mu})+\nabla^{\mu}(B^{\sigma}B^{\nu})+g^{\nu \sigma}\nabla_{\rho}(B^{\rho}B^{\mu}).
\end{align}
Here and in what follows $h_{\mu \nu }=\delta g_{\mu \nu },\,\,\,\,h^{\mu \nu }= g^{\mu \alpha } g^{\nu \beta }\delta g_{\alpha \beta }=-\delta g^{\mu \nu },\,\,\,\,h=g^{\mu \nu }\delta g_{\mu \nu }$, and $\left(d^{n-p} x\right)_{\mu_{1} \cdots \mu_{p}}=\frac{1}{p !(n-p) !} \varepsilon_{\mu_{1} \cdots \mu_{p} \nu_{p+1} \cdots \nu_{n}} d x^{\nu_{p+1}} \wedge \cdots \wedge d x^{\nu_{n}}$.

In the bumblebee gravity, the $\mathrm{U}(1)$ symmetry is broken, thus the transformation is generated by diffeomorphism $\xi$ only, i.e., $\epsilon=\xi$. The fields transformation are 
\begin{align}
	\delta_{\xi} g_{\mu \nu}&=\mathcal{L}_{\xi} g_{\mu \nu}= \nabla_{\mu} \xi_{\nu}+\nabla_{\nu} \xi_{\mu}, \\ \nonumber
	\delta_{\xi} B_{\mu}&=\mathcal{L}_{\xi} B_{\mu}=\xi^{\nu}\nabla_{\nu} B_{\mu}+B_{\nu}\nabla_{\mu}\xi^{\nu},
\end{align}
where $\mathcal{L}_{\xi} $ is the Lie derivative along the vector field $\xi$. From Eq.(\ref{NW charge density}) we can get the Noether-Wald charge density
\begin{equation}\label{key}
	\mathbf{Q}_{\xi}=\mathrm{Q}^{\mu \nu}_{\xi}\sqrt{-g}(d^{n-2}x)_{\mu \nu} ,
\end{equation}
where
\begin{align}
	\mathrm{Q}^{\mu \nu}_{\xi}=&\frac{1}{2 \kappa}\left[-2 \nabla^{[\mu} \xi^{\nu]} \right] \nonumber\\ +& \frac{\rho}{2 \kappa}\left[ 2 \nabla^{[\mu}(B^{\nu]} B_{\sigma}) \xi^{\sigma}+4\nabla_{\sigma}(B^{\sigma} B^{[\mu}) \xi^{\nu]}-\nabla_{\sigma}(2B^{\sigma} 
	B^{[\mu}  \xi^{\nu]}) \right] \nonumber\\
	-& B^{\mu \nu}B_{\sigma}\xi^{\sigma}.
\end{align}

Thus, by using Eq.(\ref{k expression}) we get the surface charge density
\begin{equation}\label{surface charge expression}
	\bm{k}_{\xi}(\delta \Phi,\Phi)=\sqrt{-g}k^{\mu \nu}_{\xi}(d^{n-2}x)_{\mu \nu} \, ,
\end{equation}
where
\begin{equation}\label{key}
	k^{\mu \nu}_{\xi}(\delta\Phi,\Phi)=k^{\mu \nu}_{g}(\delta\Phi,\Phi)+k^{\mu \nu}_{g B}(\delta\Phi,\Phi)+k^{\mu \nu}_{B}(\delta\Phi,\Phi),
\end{equation}
with
\begin{align}
	k^{\mu \nu}_{g}(\delta\Phi,\Phi)=& \frac{1}{2 \kappa}\cdot 2\left[-\frac{1}{2} h \nabla^{[ \mu} \xi^{\nu ]}+h^{\sigma [\mu} \nabla_{\sigma} \xi^{\nu]}- \xi_{\sigma} \nabla^{[\mu} h^{\nu] \sigma}-\xi^{[\mu} \nabla^{\nu]} h+\xi^{[\mu} \nabla_{\sigma} h^{\nu] \sigma} \right] , \nonumber \\
	k^{\mu \nu}_{g B}(\delta\Phi,\Phi)=&\frac{\rho}{2 \kappa}\cdot 2 \bigg[\frac{1}{2} h \nabla^{[\mu} (B^{\nu] } B_{\sigma}) \xi^{\sigma}+h \nabla_{\sigma} (B^{\sigma} B^{[\mu}) \xi^{\nu]} -\frac{1}{2} h \nabla_{\sigma}(B^{\sigma} B^{[ \mu } \xi^{\nu ] } ) \nonumber\\
	&-h^{\alpha [ \mu } \nabla_{\alpha} (B^{\nu ] } B_{\sigma}) \xi^{\sigma} + g^{\alpha [ \mu} \delta[\nabla_{\alpha} (B^{\nu] } B_{\sigma})] \xi^{\sigma}+2 \delta [\nabla_{\sigma} (B^{\sigma} B^{[ \mu})] \xi^{\nu ] }   \nonumber\\
	&-\delta[\nabla_{\sigma} (B^{\sigma} B^{[ \mu } \xi^{\nu]})] - [\nabla_{\alpha} (
	\Omega^{[ \mu | \alpha \sigma \rho|} h_{\sigma \rho}) +\Sigma^{[ \mu | \alpha \sigma|} h_{\alpha \sigma}] \cdot \xi^{\nu] } \bigg], \nonumber\\
	k^{\mu \nu}_{B}(\delta\Phi,\Phi)=& 2\left[-\frac{1}{4} h B^{\mu \nu} B_{\sigma} \xi^{\sigma}-\frac{1}{2} \delta B^{\mu \nu} B_{\sigma} \xi^{\sigma}-\frac{1}{2}B^{\mu \nu} \delta B_{\sigma} \xi^{\sigma}-B^{\sigma [ \mu} \delta B_{\sigma} \xi^{\nu] }  \right] ,
\end{align}
in which
\begin{align}\label{key35}
	&\delta B^{\mu \nu}=2h_{\alpha}{}^{[\mu} B^{\nu] \alpha}+2 g^{\alpha [ \mu } g^{\nu ] \beta} \nabla_{\alpha}[\delta B_{\beta}], \nonumber\\
	&\delta[\nabla_{\alpha} (B^{\nu} B_{\sigma})]=-B_{\sigma}h^{\nu}{}_{\beta}\nabla_{\alpha}B^{\beta}+\delta B_{\sigma}\nabla_{\alpha}B^{\nu}+\delta B^{\nu}\nabla_{\alpha}B_{\sigma}-B^{\beta}h^{\nu}{}_{\beta}\nabla_{\alpha}B_{\sigma}+B_{\sigma}\nabla_{\alpha}\delta B^{\nu}   \nonumber\\
	& + B^{\nu}\nabla_{\alpha}\delta B_{\sigma}-B^{\nu}B_{\beta}\delta \Gamma^{\beta}{}_{\alpha \sigma}-g^{\nu \rho}B_{\sigma}B_{\beta}\delta \Gamma^{\beta}{}_{\alpha \rho},  \nonumber\\
	&\delta[\nabla_{\sigma} (B^{\sigma} B^{\mu})]=2B^{(\alpha}\nabla_{\alpha}\delta B^{\mu)}+2\delta B^{(\alpha}\nabla_{\alpha}B^{\mu)}-2B^{\alpha}h^{\mu}{}_{(\alpha}\nabla_{\beta)}B^{\beta}-2h_{\alpha}{}^{\beta}B^{(\alpha}\nabla_{\beta}B^{\mu)} \nonumber\\
	&- B_{\alpha}B^{\mu}\nabla_{\beta}h^{\alpha \beta}+\frac{1}{2}B^{\alpha}B^{\mu}\nabla_{\alpha}h-\frac{1}{2}B^{\alpha}B^{\beta}\nabla^{\mu}h_{\alpha \beta} ,  \nonumber\\
	&\delta[\nabla_{\sigma} (B^{\sigma} B^{[ \mu} \xi^{\nu ]})]=-B^{\alpha}h_{\beta}{}^{[\mu}\xi^{\nu]}\nabla_{\alpha}B^{\beta}+\delta B^{\alpha}\xi^{[\nu} \nabla_{\alpha}B^{\mu]}+ B^{\alpha}\xi^{[\nu} \nabla_{\alpha}\delta B^{\mu]}+\frac{1}{2}B^{\alpha}B^{[\mu}\xi^{\nu]}\nabla_{\alpha}h \nonumber\\
	&+\frac{1}{2}B^{\alpha}\xi^{\beta}B^{[ \mu}\nabla_{\alpha}h^{\nu]}{}_{\beta}+\delta B^{\alpha}B^{[\mu}\nabla_{\alpha}\xi^{\nu]}+B^{\alpha}\delta B^{[\mu}\nabla_{\alpha}\xi^{\nu]}-B^{\alpha}B^{\beta}h^{[\mu}{}_{\beta}\nabla_{\alpha}\xi^{\nu]}+\delta B^{[\mu}\xi^{\nu]}\nabla_{\alpha}B^{\alpha} \nonumber\\
	&-B^{\alpha}h^{[\mu}{}_{\alpha}\xi^{\nu]}\nabla_{\beta}B^{\beta}-B^{\alpha}B^{[\mu}\xi^{\nu]}\nabla^{\beta}h_{\alpha \beta}+\frac{1}{2}B^{\alpha}\xi^{\beta}B^{[\mu}\nabla_{\beta}h^{\nu]}{}_{\alpha}-h_{\alpha \beta}B^{[\mu}\xi^{\nu]}\nabla^{\beta}B^{\alpha}+B^{[\mu}\xi^{\nu]}\nabla_{\alpha}\delta B^{\alpha} \nonumber\\
	&-h_{\alpha}{}^{\beta}B^{\alpha}\xi^{[\nu}\nabla_{\beta}B^{\mu]}-h_{\alpha}{}^{\beta}B^{\alpha}B^{[\mu}\nabla_{\beta}\xi^{\nu]}-\frac{1}{2}B^{\alpha}B^{\beta}\xi^{[\nu}\nabla^{\mu]}h_{\alpha \beta}-\frac{1}{2}B^{\alpha}\xi^{\beta}B^{[\mu}\nabla^{\nu]}h_{\alpha \beta}.
\end{align}

\section{Conserved charges for BTZ-like bumblebee black hole}

When the potential $V=\frac{\lambda}{2} X $ in the Lagrangian, where $\lambda$ is a Lagrange-multiplier, and there is no bumblebee field frosted at its vacuum expectation value, i.e., $B_{\mu}=b_{\mu}$, the $3$-dimensional rotating BTZ-like bumblebee black hole solution is given by \cite{Ding:2023niy}
\begin{align}\label{bmetric}
	ds^2=& -f(r) dt^2+\frac{(1+s)}{f(r)}dr^2+r^2(d\phi-\frac{j}{2r^2}dt)^2 ,\nonumber\\
	f(r)=&-\mu-\Lambda r^2 +\frac{j^2}{4r^2}, \nonumber\\
	B_{\mu}=& (0, ~B_{r},~0), \nonumber\\
	B_{r}=& b_{0} \left[ \frac{(1+s)}{f(r)} \right]^{\frac{1}{2}},
\end{align}
where $\mu$ and $j$ are mass and angular momentum parameters, respectively. $s=\rho b^2_0$ represents the spontaneous Lorentz symmetry breaking parameter, with $b_0$ being a positive constant. The negative cosmological constant $\Lambda=-\frac{1}{l^2}$ is related to the Lagrange-multiplier $\lambda$, $\Lambda=(1+s)\frac{\kappa \lambda}{2 \rho}$.

This solution has two horizons: inner horizon $r_{-}$ and out horizon $r_{+}$, which solve $f(r_{\pm})=0$,
\begin{equation}
	r_{\pm}=\frac{\sqrt{2} }{2} l \sqrt{\mu \pm \sqrt{\mu^2-\frac{j^2}{l^2}}}.
\end{equation}

Since the integration in Eq.(\ref{dhintgration}) is independent of the choice of integration surface $\partial \Sigma$, we take $\partial \Sigma$ to be the circle $S$ of constant $(t,r)$ for simplicity and take the limit $r \rightarrow \infty$. Then, the conserved charge variation in $3$-dimensional gravity associated with exact symmetry $\eta$ can be 
\begin{equation}\label{deltaH}
	\hat{\delta} H_{\eta}=\oint_{S} \bm{k}_{\eta}(\hat{\delta} \Phi, \hat{\Phi})=\int_{0}^{2 \pi} \lim_{r\to \infty}\sqrt{-\hat{g}} \, k_{\eta}^{t r}(\hat{\delta} \Phi, \hat{\Phi}) \mathrm{d} \phi,
\end{equation}
where $k_{\eta}^{t r}$ is the $tr$ component of $k_{\eta}^{\mu \nu}$. The dynamical fields are $\hat\Phi=(\hat g_{\mu \nu}, \hat B_{\mu})$, parametrized by $p_{\alpha}=\{\mu, j \}$. The parametric variations are given by
\begin{align}\label{parametric variations}
	\hat{\delta} g_{\mu \nu}=&\frac{\partial \hat{g}_{\mu \nu}}{\partial \mu} \delta \mu+\frac{\partial \hat{g}_{\mu \nu}}{\partial j} \delta j ,\nonumber \\
	\hat{\delta} B_{\mu}=&\frac{\partial \hat{B}_{\mu}}{\partial \mu} \delta \mu+\frac{\partial \hat{B}_{\mu}}{\partial j} \delta j .
\end{align}
The stationary axial symmetric black hole metric has two killing vectors $\xi=\partial_{t}$ and $\xi=\partial_{\phi}$, which are exact symmetries of the spacetime, and any linear combination of these two with $\mu, j$-dependent coefficients is also a killing. \\
\textbf{Mass:} By choosing the exact symmetry $\eta_{{}_{M}}= \partial_{t}$ and substituting the surface charge expressions (\ref{surface charge expression})-(\ref{key35}) along with the parametric variations (\ref{parametric variations}) into Eq.(\ref{deltaH}), we can obtain the total mass
\begin{equation}\label{key}
	\hat{\delta} M=\hat{\delta} H_{\eta_{{}_{\mathrm{M}}}}=\frac{\sqrt{1+s} \delta \mu}{4 G}  \quad \Rightarrow \quad M=\frac{\sqrt{1+s} \mu}{4 G} .
\end{equation}
\textbf{Angular momentum:} By choosing the exact symmetry $\eta_{{}_{J}}=-\partial_{\phi}$, and using Eq.(\ref{deltaH}), the angular momentum is given by
\begin{equation}\label{key}
	\hat{\delta} J=\hat{\delta} H_{\eta_{{}_{\mathrm{J}}}}=\frac{\sqrt{1+s} \delta j}{4 G}  \quad \Rightarrow \quad J=\frac{\sqrt{1+s} j}{4 G} .
\end{equation}
\textbf{Entropy:} The killing vector generating the killing horizon of the black hole is
\begin{equation}\label{key}
	\zeta=\partial_{t}+\Omega_{\mathrm{H}}\partial_{\phi}.
\end{equation}
The surface gravity, temperature and angular velocity on the horizon are given by \cite{Compere:2018aar}
\begin{align}
	\kappa_{{}_{\mathrm{H}}}=&\sqrt{-\frac{1}{2}(\nabla_{\mu}\zeta_{\nu})(\nabla^{\mu}\zeta^{\nu})}|_{r_{{}_{\mathrm{H}}}}= \frac{r^2_{\pm}-r^2_{\mp}}{\sqrt{1+s} l^2 r_{\pm}},  \nonumber\\
	T_{\mathrm{H}}=& \frac{\kappa_{{}_{\mathrm{H}}}}{2 \pi},  \nonumber\\
	\Omega_{\mathrm{H}}=&-\frac{g_{t \varphi}}{g_{\varphi \varphi}}|_{r_{{}_{\mathrm{H}}}}=\frac{j}{2 {r^2_{{}_{\mathrm{H}}}}}.
\end{align}
where $r_{{}_{\mathrm{H}}} \equiv r_{\pm}$. By choosing the exact symmetry $\eta_{{}_{\mathrm{H}}}=\frac{2 \pi}{\kappa_{{}_{\mathrm{H}}}} \zeta $, and using Eq.(\ref{deltaH}), we obtain the entropy
\begin{equation}\label{entropy}
	\hat{\delta} S_{\mathrm{H}}=\hat{\delta} H_{\eta_{{}_{\mathrm{H}}}}=\frac{(1+s)\pi \delta r_{{}_{\mathrm{H}}}}{G} \quad \Rightarrow \quad S_{\mathrm{H}}=\frac{(1+s)\pi r_{{}_{\mathrm{H}}}}{G} .
\end{equation}
\textbf{The first law:} With the decomposition
\begin{equation}\label{key}
	\eta_{{}_{\mathrm{H}}}=\frac{1}{T_{H}}(\eta_{{}_{M}}-\Omega_{\mathrm{H}}\eta_{{}_{J}}),
\end{equation}
and the linearity of $\hat{\delta} H_{\eta_{{}_{\mathrm{H}}}}$ in $\eta$, the first law is satisfied
\begin{equation}\label{The first law}
	\delta S=\frac{1}{T_{\mathrm{H}}}(\delta M-\Omega_{\mathrm{H}} \delta J).
\end{equation}

In our formulation the reference points for the charges above are chosen to vanish when $\mu=j=0$. The total mass, angular momentum and entropy are all affected by the factor $\sqrt{1+s}$, arising due to LV effect. Our results differ from those obtained by the conventional thermodynamic method in Ref. \cite{Ding:2023niy}, where $M=\frac{\mu}{8 G}$, $J=\frac{j}{8 G}$ and $S=\frac{\sqrt{1+s} \pi r_{{}_{\mathrm{H}}}}{2 G} $, with constant G having been retrieved. On the other hand, the Bekenstein-Hawking entropy (\ref{entropy}) was straightforwardly derived as a conserved charge, and the first law is perfectly satisfied. This is an advantage of the SPSM. When $s \rightarrow 0$ , this solution reduces to the exact BTZ black hole solution, and $M=\frac{\mu}{4 G}$, $J=\frac{j}{4 G}$, $S=\frac{ \pi r_{{}_{\mathrm{H}}}} {G}$ (up to factor $\frac{1}{2}$ , which comes from the choice of the gravitational coupling constant $\kappa=4 \pi G$ in original black hole solution in Ref. \cite{Ding:2023niy} and also in our case). Although the bumblebee field strength $B_{\mu \nu}$ is analogous to the electromagnetic field strength $F_{\mu \nu}$, the $\mathrm{U(1)}$ internal gauge symmetry is broken in the bumblebee theory. As a result, there are no corresponding $\mathrm{U(1)}$ conserved charges (such as ‘electromagnetic charges $Q_{\mathrm{I}}$’) and no terms like $\Phi^\mathrm{I}_{\mathrm{H}} \delta Q_{\mathrm{I}}$ appearing in the first law. At this point, it is different from the electromagnetic theories with the $\mathrm{U(1)}$ symmetry. As a consistent check, in the next section we will derive the microscopic entropy of dual CFT by the AdS/CFT correspondence, and find that the Bekenstein-Hawking entropy (\ref{entropy}) is in precise agreement with the microscopic entropy of the dual CFT.

\section{Asymptotic symmetries}
In this section we will analysis the asymptotic symmetries in $\mathrm{AdS}_3$ background, and obtain the warped conformal algebra in the near horizon region by employing the new boundary conditions. 

\subsection{AdS/CFT correspondence}
Setting $\mu=0=j$ in the metric and bumblebee field of the BTZ-like black hole (\ref{bmetric}), we can get the $\mathrm{AdS}_3$ background spacetime
\begin{align}\label{adsbackground}
	d\bar{s}^2=& -\frac{r^2}{l^2}dt^2+\frac{(1+s)l^2}{r^2}dr^2+r^2d\phi^2, \nonumber\\
	\bar{B}_{\mu}=& \left(0, ~\frac{b_0 \sqrt{1+s} l}{r}, ~0\right).
\end{align}

By using the Brown-Henneaux boundary conditions \cite{Strominger:1997eq}
\begin{equation}\label{dm}
	\delta g_{\mu \nu}=\mathcal{O}
	\left(\begin{array}{ccc}
		1~~~ & \frac{1}{r^3}~~~ & 1  \\
		& \frac{1}{r^4}~~~ & \frac{1}{r^3} \\
		&               & 1
	\end{array}\right),
\end{equation}
%
and choose the following boundary conditions for $B_{\mu}$,
\begin{equation}\label{dB}
	\delta B_{\mu}=\mathcal{O} \left(1, ~\frac{1}{r^3}, ~1\right),
\end{equation}
one can get the asymptotic killing vectors $\xi=\xi^{\mu} \partial_{\mu}$, which preserve (\ref{dm}) and (\ref{dB}), 
\begin{align}\label{ak}
	\xi^{t}=& l(\mathrm{T}^{+}+\mathrm{T}^{-})+\frac{l^3}{2 r^2}(\partial^2_{+}\mathrm{T}^{+}+\partial^2_{-}\mathrm{T}^{-})+\mathcal{O}(\frac{1}{r^4}) \nonumber\\
	\xi^{\phi}=& \mathrm{T}^{+}-\mathrm{T}^{-}-\frac{l^2}{2 r^2}(\partial^2_{+}\mathrm{T}^{+}-\partial^2_{-}\mathrm{T}^{-})+\mathcal{O}(\frac{1}{r^4}) \nonumber\\
	\xi^{r}=&-r(\partial_{+}\mathrm{T}^{+}+\partial_{-}\mathrm{T}^{-})+\mathcal{O}(\frac{1}{r}) ,
\end{align}
where $x^{\pm}=\frac{t}{l} \pm \phi$, $2\partial_{\pm}=l\partial_{t} \pm \partial_{\phi}$, and $\mathrm{T}^{\pm}=\mathrm{T}^{\pm}(x^{\pm}) $. By taking modes $\xi^{\pm}_{m}=\xi(\mathrm{T}^{\pm}= \frac{1}{2}e^{i m x^{\pm}})$, $m\in \mathbb{Z}$, one can get the Witt algebras through Lie brackets,
\begin{align}
	i\left[\xi^{\pm}_{m} , \xi^{\pm}_{n} \right]=& (m-n)\xi^{\pm}_{(m+n)}, \nonumber\\
	i\left[\xi^{+}_{m} , \xi^{-}_{n} \right]=& 0.
\end{align}

For the generator $\xi^{+}_{m}$, the Poisson bracket algebra (\ref{poissonb}) becomes
\begin{equation}\label{key}
	i\left\{ \tilde{L}^{+}_m, \tilde{L}^{+}_n \right\}=(m-n)\tilde{L}^{+}_{m+n}-i H_{ [ \xi^{+}_{m},\xi^{+}_{n} ] } (\bar{\Phi}) +i \oint_{S}\bm {k}_{\xi^{+}_{m}}(\delta_{\xi^{+}_{n}} \bar{\Phi},\bar{\Phi}),
\end{equation}
where the central extention term
\begin{equation}\label{key}
	i \oint_{S}\bm {k}_{\xi^{+}_{m}}(\delta_{\xi^{+}_{n}} \bar{\Phi},\bar{\Phi})=\frac{1}{12} \frac{3 \sqrt{1+s} l}{G}m^3\delta_{m+n,0}.
\end{equation}
Similarly, for $i\left\{ \tilde{L}^{-}_m, \tilde{L}^{-}_n \right\}$, the central extention term
\begin{equation}\label{key}
	i \oint_{S}\bm {k}_{\xi^{-}_{m}}(\delta_{\xi^{-}_{n}} \bar{\Phi},\bar{\Phi})=\frac{1}{12} \frac{3 \sqrt{1+s} l}{G}m^3\delta_{m+n,0},
\end{equation}
where $\tilde{L}^{+}_m \equiv H_{\xi^{+}_m}$, $\tilde{L}^{-}_m \equiv H_{\xi^{-}_m}$, and $\bar {\Phi}$ is the background (\ref{adsbackground}). To contain centerless subalgebra for $m=-1, 0, +1$, we shift the zero modes of the charges
\begin{align}
	L^{+}_m=&\tilde{L}^{+}_m + \delta_{m,0} \frac{3 \sqrt{1+s} l}{24G}, \nonumber\\
	L^{-}_n=&\tilde{L}^{-}_n + \delta_{n,0} \frac{3 \sqrt{1+s} l}{24G}.
\end{align}

Then, the Poisson bracket algebra of the charges are two copies of the Virasoro algebra
\begin{align}
	i\left\{ L^{+}_m, L^{+}_n \right\}=&(m-n) L^{+}_{m+n}+\frac{c}{12} m(m^2 -1)\delta_{m+n,0}, \nonumber\\
	i\left\{ L^{+}_m, L^{-}_n \right\}=&0, \nonumber\\
	i\left\{ L^{-}_m, L^{-}_n \right\}=&(m-n) L^{-}_{m+n}+\frac{c}{12} m(m^2 -1)\delta_{m+n,0}.
\end{align}
The two central charges are the same, and with
\begin{equation}\label{key}
	c=\frac{3 \sqrt{1+s} l}{G}.
\end{equation}

By using the Cardy formula \cite{Cardy:1986ie,deBoer:2010ac}, and modified by the LV factor $\sqrt{1+s}$, the microscopic entropy of the dual CFT is
\begin{equation}\label{key}
	S_{CFT}=2\pi (\sqrt{1+s}) \sqrt{\frac{c}{6}(L^{-}_0 -  \frac{c}{24})} \pm 2\pi (\sqrt{1+s}) \sqrt{\frac{c}{6}(L^{+}_0 - \frac{c}{24})},
\end{equation}
The zero modes of $\tilde{L}^{+}_m$ and $\tilde{L}^{-}_n$ for the BTZ-like black hole (\ref{bmetric}) are
\begin{align}
	\tilde{L}^{+}_0=& H_{\xi^{+}_{0}}=\frac{\sqrt{1+s} l }{8 G} \left( \mu-\frac{ j}{l} \right), \nonumber\\
	\tilde{L}^{-}_0=& H_{\xi^{-}_{0}}=\frac{\sqrt{1+s} l }{8 G} \left( \mu+\frac{ j}{l} \right).
\end{align}
Thus, we can obtain 
\begin{equation}\label{key}
	S_{CFT}=\frac{(1+s) \pi r_{{}_{\mathrm{H}}}}{G}.
\end{equation}
This result agrees precisely with the Bekenstein-Hawking entropy (\ref{entropy}). 


\subsection{Warped conformal symmetries in near horizon region}
In this subsection we will use the Detournay-Smoes-Wutte boundary conditions \cite{Detournay:2023zni} to study the asymptotic symmetries of the near horizon extremal BTZ-like black hole.

In the extremal case, the BTZ-like metric and the bumblebee field (\ref{bmetric}) are
\begin{align}
	ds^2=& -\frac{(r^2-r^2_{h})^2 }{l^2 r^2} dt^2+\frac{(1+s) l^2 r^2}{(r^2-r^2_{h})^2}dr^2+r^2(d\phi-\frac{r^2_{h}}{l r^2} dt)^2, \nonumber\\
	B_{\mu}=& \left(0,~b_0 \bigg[\frac{(1+s) l^2  r^2}{(r^2-r^2_{h})^2} \bigg]^{\frac{1}{2}},~0 \right),
\end{align}
where $r_h=r_{+}=r_{-}=\frac{\sqrt2}{2} l \sqrt{\mu}$. We consider the change of coordinates
\begin{equation}\label{key}
	t=\frac{\tau}{\varepsilon}, \quad r^2=r^2_h+\varepsilon \rho, \quad \phi=\varphi+\frac{\tau}{\varepsilon l}.
\end{equation}
By taking limit $\varepsilon \rightarrow 0$, we can obtain the near horizon geometry of the extremal BTZ-like metric and bumblebee field
\begin{align}\label{NHEG}
	ds^2=& \frac{(1+s) l^2 }{4 \rho^2} d\rho^2+\frac{2 \rho}{l}d\tau d\varphi+r^2_h d\varphi^2, \nonumber\\
	B_{\mu}=& \left(0,~\frac{b_0 \sqrt{1+s} l}{2\rho}, ~0 \right).
\end{align}

We now impose the Detournay-Smoes-Wutte boundary conditions in Ref. \cite{Detournay:2023zni} on the near horizon geometry of the extremal BTZ-like metric and bumblebee field (\ref{NHEG}), which is a higher dimensional uplift of Godet-Marteau boundary conditions in two-dimensional gravity \cite{Godet:2020xpk}, and it is given by a finite coordinate transformation
\begin{equation}\label{ctransformations}
	\tau \rightarrow \mathcal{F}(\tau), \quad \rho \rightarrow \frac{1}{\mathcal{F'}}( \rho+\mathcal{G'}(\tau) ), \quad \varphi \rightarrow \varphi-\mathcal{G}(\tau).
\end{equation}
Here $\mathcal{F}(\tau)$ and $\mathcal{G}(\tau)$ are arbitrary functions of $\tau$. This transformation yields the metric and bumblebee field components
\begin{align}
	g_{\tau \tau}=& r^2_h \mathcal{G'}(\tau)^2-\frac{2}{l} \mathcal{G'}(\tau) (\rho+\mathcal{G'}(\tau))+\frac{(1+s) l^2 (\mathcal{G''}(\tau)-\mathcal{H}(\tau)(\rho+\mathcal{G'}(\tau)))^2}{4 (\rho+\mathcal{G'}(\tau))^2 }, \nonumber\\
	g_{\tau \rho}=& \frac{(1+s) l^2 (\mathcal{G''}(\tau)-\mathcal{H}(\tau)(\rho+\mathcal{G'}(\tau))) }{4 (\rho+\mathcal{G'}(\tau))^2 }, \nonumber\\
	g_{\tau \varphi}=& -r^2_h \mathcal{G'}(\tau)+\frac{\rho+\mathcal{G'}(\tau)}{l}, \nonumber\\
	g_{\rho \rho}=& \frac{(1+s) l^2 }{4 (\rho+\mathcal{G'}(\tau))^2 }, \quad  g_{\rho \varphi}=0, \quad  g_{\varphi \varphi}= r^2_h, \nonumber\\
	B_{\tau}=&- \frac{b_0 \sqrt{1+s} l (\rho \mathcal{H}(\tau)+\mathcal{H}(\tau)\mathcal{G'}(\tau)-\mathcal{G''}(\tau) ) }{2 (\rho+\mathcal{G'}(\tau)) }, \nonumber\\
	B_{\rho}=& \frac{b_0 \sqrt{1+s} l }{2 (\rho+\mathcal{G'}(\tau)) }, \nonumber\\
	B_{\varphi}=& 0,
\end{align}
where $\mathcal{H}(\tau) \equiv \mathcal{F''}(\tau)/\mathcal{F'}(\tau) $. The near horizon geometry (\ref{NHEG}) is restored when $\mathcal{G'}(\tau)=0=\mathcal{H}(\tau)$.

The asymptotic killing vectors generating the transformations (\ref{ctransformations}) are given by
\begin{equation}\label{akilling}
	\xi=\epsilon(\tau) \partial_{\tau}-(\rho \epsilon'(\tau)+\zeta'(\tau) )\partial_{\rho}+\zeta(\tau)\partial_{\varphi},
\end{equation}
where $\epsilon(\tau)$ and $\zeta(\tau)$ are two arbitrary functions of $\tau$. By applying the Lie derivative on the metric and bumblebee field, the variations of $\mathcal{G'}(\tau)$ and $\mathcal{H}(\tau)$ are given by
\begin{align}
	\delta_{\xi }\mathcal{G'}(\tau)=& \epsilon'(\tau) \mathcal{G'}(\tau)+\epsilon(\tau) \mathcal{G''}(\tau)-\zeta'(\tau), \nonumber\\
	\delta_{\xi }\mathcal{H}(\tau)=& \epsilon'(\tau) \mathcal{H}(\tau)+\epsilon''(\tau)+\epsilon(\tau) \mathcal{H'}(\tau).
\end{align}
To obtain the symmetry algebra, we assume that $\tau$ is periodic with period $L \in i\mathbb{R}$, where $i$ is the imaginary unit. The asymptotic killing vector (\ref{akilling}) can be divided into modes
\begin{align}
	l_{n}=& \xi \left(\epsilon(\tau)=\frac{L}{2 \pi} e^{2\pi i n \tau/L},~\zeta(\tau)=0 \right), \nonumber\\
	j_{n}=& \xi \left(\epsilon(\tau)=0,~\zeta(\tau)=\frac{L}{2 \pi i} e^{2\pi i n \tau/L} \right),
\end{align}
where $n \in \mathbb{Z}$. Under the Lie bracket, these modes satisfy the warped Witt algebra
\begin{align}
	i \left[l_{m}, l_{n} \right]=& (m-n) l_{m+n}, \nonumber\\
	i \left[l_{m}, j_{n} \right]=& -n j_{m+n}, \nonumber\\
	i \left[j_{m}, j_{n} \right]=& 0. 
\end{align}

To obtain finite and non-zero charges (\ref{eqdh}), we take $\partial{\Sigma}$ to be the circle S of constant $(r,\varphi)$ and take the limit $r \rightarrow \infty$. By taking variations of the metric and bumblebee field
\begin{align}
	\delta g_{\mu \nu}=& \frac{\partial g_{\mu \nu}}{\partial \mathcal{G'}(\tau) } \delta \mathcal{G'}(\tau)+\frac{\partial g_{\mu \nu}}{\partial \mathcal{G''}(\tau) } \delta \mathcal{G''}(\tau)+\frac{\partial g_{\mu \nu}}{\partial \mathcal{H}(\tau) } \delta \mathcal{H}(\tau),  \nonumber\\
	\delta B_{\mu}=& \frac{\partial B_{\mu}}{\partial \mathcal{G'}(\tau) } \delta \mathcal{G'}(\tau)+\frac{\partial B_{\mu}}{\partial \mathcal{G''}(\tau) } \delta \mathcal{G''}(\tau)+\frac{\partial B_{\mu }}{\partial \mathcal{H}(\tau) } \delta \mathcal{H}(\tau), 
\end{align}
the corresponding charge variation (\ref{eqdh}) associated with $\xi$ can be given by 
\begin{equation}
	\delta H_{\xi}=\frac{\sqrt{1+s} l}{16 \pi G} \int^{L}_0 d\tau \left(4 \mu \zeta(\tau) \delta \mathcal{G'}(\tau)-2\mu \epsilon(\tau) \delta \mathcal{G'}(\tau)^2 +\epsilon'(\tau) \delta \mathcal{H}(\tau)+\epsilon(\tau) \delta \mathcal{H}(\tau)^2  \right),
\end{equation}
which is integrable. Under the background (\ref{NHEG}), the charges are given by
\begin{align}
	\hat{L}_{n}=&H_{l_{n}}=\frac{\sqrt{1+s} l}{16 \pi G} \int^{L}_0 d\tau \left(- 2\mu \mathcal{G'}(\tau)^2+\frac{4 \pi i n}{L} \mathcal{H}(\tau)+ \mathcal{H}(\tau)^2  \right)\frac{L}{2 \pi} e^{2\pi i n \tau/L}, \nonumber\\
	\hat{J}_{n}=&H_{j_{n}}=\frac{\sqrt{1+s} l}{16 \pi G} \int^{L}_0 d\tau \left(4\mu \mathcal{G'}(\tau) \right)\frac{L}{2 \pi i} e^{2\pi i n \tau/L}.
\end{align}

The Poisson bracket algebra (\ref{poissonb}) of $\hat{L}_{n}$ and $\hat{J}_{n}$ is a Virasoro-Kac-Moody $\mathrm{U(1)}$ algebra, the symmetry algebra of a WCFT,
\begin{align}
	i \left\{ \hat{L}_{m}, \hat{L}_{n} \right\}=& (m-n) \hat{L}_{m+n}+\frac{\hat c}{12} m^3 \delta_{m+n,0}, \nonumber\\
	i \left\{ \hat{L}_{m}, \hat{J}_{n} \right\}=& -n \hat{J}_{m+n}-i\hat{\kappa} m^2 \delta_{m+n,0}, \nonumber\\
	i \left\{ \hat{J}_{m}, \hat{J}_{n} \right\}=& \frac{\hat k}{2} m \delta_{m+n,0}, 
\end{align}
with central charges
\begin{equation}\label{key}
	\hat c=\frac{3 \sqrt{1+s} l}{G}, \quad \quad \hat{\kappa}=0, \quad \quad \hat k=\frac{ \sqrt{1+s} l L^2 \mu}{4 \pi^2 G}.
\end{equation}

The $\hat c$ is the same as the Virasoro central charge $c$ of the AdS/CFT correspondence.

\section{Conclusions and Outlooks}
In this paper, we employ the SPSM to investigate the conserved charges of a rotating BTZ-like black hole in Einstein-bumblebee theory of gravity. We explicitly computed the black hole mass, angular momentum and entropy, demonstrating that these quantities are influenced by LV parameter. Our results differ from those obtained in Ref. \cite{Ding:2023niy} using the conventional thermodynamic method. Additionally, we derived the first law of black hole thermodynamics within this framework. As an asymptotical AdS spacetime, we studied the AdS/CFT correspondence and obtained the asymptotic charge algebra, which are two copies of the Virasoro algebra with non-trivial central charges. By applying the Cardy formula, we computed the microscopic entropy of the dual CFT, and found it is in exact agreement with the Bekenstein-Hawking entropy calculated by SPSM. Thus, we confirmed our results for the conserved charges through entropy matching and by verifying the consistency of the first law of black hole thermodynamics. Furthermore, in the near horizon region, we used the Detournay-Smoes-Wutte boundary conditions to study the near horizon geometry of extremal BTZ-like black hole. We obtained a Virasoro-Kac-Moody $\mathrm{U(1)}$ algebra, which is the symmetry algebra of WCFT. And the central charge $\hat c$ of the Virasoro-Kac-Moody $\mathrm{U(1)}$ algebra is found to be identical to the central charge $c$ of the Virasoro algebra in AdS/CFT correspondence.

Evidently, the conserved charges and central charges are affected by the LV parameter. We speculate that similar modifications arise in other bumblebee black hole solutions \cite{Casana:2017jkc,Maluf:2020kgf,Jha:2020pvk,Filho:2022yrk,Xu:2022frb,Liu:2024axg}, which may also affect the thermodynamic properties. We expect this deviation from General Relativity, as well as the LV parameter for four-dimensional bumblebee gravity, can be further studied and constrained by using astrophysical data, including solar system test \cite{Casana:2017jkc}, Event Horizon Telescope (EHT) observables of black hole shadows \cite{Wang:2021irh,Islam:2024sph,Ali:2024ssf,Pantig:2025bpd}, quasi-periodic oscillations (QPO) \cite{Wang:2021gtd}, and black hole X-ray data \cite{Gu:2022grg}. On the other hand, our study provides a first step towards building a holographic dual for near horizon geometry of bumblebee black holes. Through the Virasoro-Kac-Moody $\mathrm{U(1)}$ algebra, as well as the corresponding central charges and Kac-Moody level, many properties can be further explored, for example, deriving a Cardy-type formula in WCFT, exploring holographic thermodynamics, studing holographic entanglement entropy, phase transitions, complex, bulk reconstruction, calculating correlation functions and studing anomalies. Furthermore, the enhanced ASG of $\mathrm{WAdS_3}$ spacetimes consist of a Virasoro-Kac-Moody $\mathrm{U(1)}$ algebra, which inspires us to search for exact $\mathrm{WAdS_3}$ and warped black hole solutions in the bumblebee gravity. These are potential future research directions along this line.


~\\
\noindent \textbf{Acknowledgements}\\ \\
The author would like to thank Dr. Mingzhi Wang and Dr. Feiyi Liu for helpful discussions, and especially grateful to Prof. Xiang-Hua Zhai for carefully reading and improving this manuscript.


{\small 
	\begin{thebibliography}{10}


\bibitem{Noether:1918zz}
E.~Noether,
``Invariant Variation Problems,''
Gott. Nachr. \textbf{1918} (1918), 235-257
[arXiv:physics/0503066 [physics]].


\bibitem{Bondi:1962px}
H.~Bondi, M.~G.~J.~van der Burg and A.~W.~K.~Metzner,
``Gravitational waves in general relativity. 7. Waves from axisymmetric isolated systems,''
Proc. Roy. Soc. Lond. A \textbf{269} (1962), 21-52


\bibitem{Sachs:1962wk}
R.~K.~Sachs,
``Gravitational waves in general relativity. 8. Waves in asymptotically flat space-times,''
Proc. Roy. Soc. Lond. A \textbf{270} (1962), 103-126


\bibitem{Brown:1986nw}
J.~D.~Brown and M.~Henneaux,
``Central Charges in the Canonical Realization of Asymptotic Symmetries: An Example from Three-Dimensional Gravity,''
Commun. Math. Phys. \textbf{104} (1986), 207-226


\bibitem{Maldacena:1998bw}
J.~M.~Maldacena and A.~Strominger,
``AdS(3) black holes and a stringy exclusion principle,''
JHEP \textbf{12} (1998), 005
[arXiv:hep-th/9804085 [hep-th]].


\bibitem{Banados:1992wn}
M.~Banados, C.~Teitelboim and J.~Zanelli,
``The Black hole in three-dimensional space-time,''
Phys. Rev. Lett. \textbf{69} (1992), 1849-1851
[arXiv:hep-th/9204099 [hep-th]].


\bibitem{Banados:1992gq}
M.~Banados, M.~Henneaux, C.~Teitelboim and J.~Zanelli,
``Geometry of the (2+1) black hole,''
Phys. Rev. D \textbf{48} (1993), 1506-1525
[erratum: Phys. Rev. D \textbf{88} (2013), 069902]
[arXiv:gr-qc/9302012 [gr-qc]].


\bibitem{Strominger:1997eq}
A.~Strominger,
``Black hole entropy from near horizon microstates,''
JHEP \textbf{02} (1998), 009
[arXiv:hep-th/9712251 [hep-th]].


\bibitem{Guica:2008mu}
M.~Guica, T.~Hartman, W.~Song and A.~Strominger,
``The Kerr/CFT Correspondence,''
Phys. Rev. D \textbf{80} (2009), 124008
[arXiv:0809.4266 [hep-th]].


\bibitem{Detournay:2012pc}
S.~Detournay, T.~Hartman and D.~M.~Hofman,
``Warped Conformal Field Theory,''
Phys. Rev. D \textbf{86} (2012), 124018
[arXiv:1210.0539 [hep-th]].


\bibitem{Compere:2018aar}
G.~Comp\`ere and A.~Fiorucci,
``Advanced Lectures on General Relativity,''
[arXiv:1801.07064 [hep-th]].


\bibitem{Adami:2017phg}
H.~Adami, M.~R.~Setare, T.~C.~Sisman and B.~Tekin,
``Conserved Charges in Extended Theories of Gravity,''
Phys. Rept. \textbf{834} (2019), 1
[arXiv:1710.07252 [hep-th]].


\bibitem{Frodden:2019ylc}
E.~Frodden and D.~Hidalgo,
``Surface Charges Toolkit for Gravity,''
Int. J. Mod. Phys. D \textbf{29} (2020) no.06, 2050040
[arXiv:1911.07264 [hep-th]].


\bibitem{Lee:1990nz}
J.~Lee and R.~M.~Wald,
``Local symmetries and constraints,''
J. Math. Phys. \textbf{31} (1990), 725-743


\bibitem{Wald:1993nt}
R.~M.~Wald,
``Black hole entropy is the Noether charge,''
Phys. Rev. D \textbf{48} (1993) no.8, R3427-R3431
[arXiv:gr-qc/9307038 [gr-qc]].


\bibitem{Iyer:1994ys}
V.~Iyer and R.~M.~Wald,
``Some properties of Noether charge and a proposal for dynamical black hole entropy,''
Phys. Rev. D \textbf{50} (1994), 846-864
[arXiv:gr-qc/9403028 [gr-qc]].


\bibitem{Hajian:2015xlp}
K.~Hajian and M.~M.~Sheikh-Jabbari,
``Solution Phase Space and Conserved Charges: A General Formulation for Charges Associated with Exact Symmetries,''
Phys. Rev. D \textbf{93} (2016) no.4, 044074
[arXiv:1512.05584 [hep-th]].


\bibitem{Hajian:2016kxx}
K.~Hajian,
``Conserved charges and first law of thermodynamics for Kerr\textendash{}de Sitter black holes,''
Gen. Rel. Grav. \textbf{48} (2016) no.8, 114
[arXiv:1602.05575 [gr-qc]].


\bibitem{Ghodrati:2016vvf}
M.~Ghodrati, K.~Hajian and M.~R.~Setare,
``Revisiting Conserved Charges in Higher Curvature Gravitational Theories,''
Eur. Phys. J. C \textbf{76} (2016) no.12, 701
[arXiv:1606.04353 [hep-th]].


\bibitem{Ding:2020bwa}
H.~F.~Ding and X.~H.~Zhai,
``Entropies and The First Laws of Black Hole Thermodynamics in Einstein-aether-Maxwell Theory,''
Class. Quant. Grav. \textbf{37} (2020) no.18, 185015
[arXiv:2001.06261 [gr-qc]].


\bibitem{Guo:2024zaa}
W.~Guo,
``Notes on solution phase space and BTZ black hole,''
Eur. Phys. J. C \textbf{84} (2024) no.11, 1198
[arXiv:2411.14247 [gr-qc]].


\bibitem{Mattingly:2005re}
D.~Mattingly,
``Modern tests of Lorentz invariance,''
Living Rev. Rel. \textbf{8} (2005), 5
[arXiv:gr-qc/0502097 [gr-qc]].


\bibitem{Horava:2009uw}
P.~Horava,
``Quantum Gravity at a Lifshitz Point,''
Phys. Rev. D \textbf{79} (2009), 084008
[arXiv:0901.3775 [hep-th]].


\bibitem{Arkani-Hamed:2003pdi}
N.~Arkani-Hamed, H.~C.~Cheng, M.~A.~Luty and S.~Mukohyama,
``Ghost condensation and a consistent infrared modification of gravity,''
JHEP \textbf{05} (2004), 074
[arXiv:hep-th/0312099 [hep-th]].


\bibitem{Frey:2003jq}
A.~R.~Frey,
``String theoretic bounds on Lorentz violating warped compactification,''
JHEP \textbf{04} (2003), 012
[arXiv:hep-th/0301189 [hep-th]].


\bibitem{Cline:2003xy}
J.~M.~Cline and L.~Valcarcel,
``Asymmetrically warped compactifications and gravitational Lorentz violation,''
JHEP \textbf{03} (2004), 032
[arXiv:hep-ph/0312245 [hep-ph]].


\bibitem{Jacobson:2000xp}
T.~Jacobson and D.~Mattingly,
``Gravity with a dynamical preferred frame,''
Phys. Rev. D \textbf{64} (2001), 024028
[arXiv:gr-qc/0007031 [gr-qc]].


\bibitem{Eling:2004dk}
C.~Eling, T.~Jacobson and D.~Mattingly,
``Einstein-Aether theory,''
[arXiv:gr-qc/0410001 [gr-qc]].


\bibitem{Jacobson:2007veq}
T.~Jacobson,
``Einstein-aether gravity: A Status report,''
PoS \textbf{QG-PH} (2007), 020
[arXiv:0801.1547 [gr-qc]].


\bibitem{Kostelecky:1988zi}
V.~A.~Kostelecky and S.~Samuel,
``Spontaneous Breaking of Lorentz Symmetry in String Theory,''
Phys. Rev. D \textbf{39} (1989), 683


\bibitem{Kostelecky:1989jw}
V.~A.~Kostelecky and S.~Samuel,
``Gravitational Phenomenology in Higher Dimensional Theories and Strings,''
Phys. Rev. D \textbf{40} (1989), 1886-1903


\bibitem{Kostelecky:2003fs}
V.~A.~Kostelecky,
``Gravity, Lorentz violation, and the standard model,''
Phys. Rev. D \textbf{69} (2004), 105009
[arXiv:hep-th/0312310 [hep-th]].


\bibitem{Kostelecky:2010ze}
A.~V.~Kostelecky and J.~D.~Tasson,
``Matter-gravity couplings and Lorentz violation,''
Phys. Rev. D \textbf{83} (2011), 016013
[arXiv:1006.4106 [gr-qc]].


\bibitem{Casana:2017jkc}
R.~Casana, A.~Cavalcante, F.~P.~Poulis and E.~B.~Santos,
``Exact Schwarzschild-like solution in a bumblebee gravity model,''
Phys. Rev. D \textbf{97} (2018) no.10, 104001
[arXiv:1711.02273 [gr-qc]].


\bibitem{Maluf:2020kgf}
R.~V.~Maluf and J.~C.~S.~Neves,
``Black holes with a cosmological constant in bumblebee gravity,''
Phys. Rev. D \textbf{103} (2021) no.4, 044002
[arXiv:2011.12841 [gr-qc]].


\bibitem{Ding:2019mal}
C.~Ding, C.~Liu, R.~Casana and A.~Cavalcante,
``Exact Kerr-like solution and its shadow in a gravity model with spontaneous Lorentz symmetry breaking,''
Eur. Phys. J. C \textbf{80} (2020) no.3, 178
[arXiv:1910.02674 [gr-qc]].


\bibitem{Ding:2023niy}
C.~Ding, Y.~Shi, J.~Chen, Y.~Zhou, C.~Liu and Y.~Xiao,
``Rotating BTZ-like black hole and central charges in Einstein-bumblebee gravity,''
Eur. Phys. J. C \textbf{83} (2023) no.7, 573
[arXiv:2302.01580 [gr-qc]].


\bibitem{Santos:2014nxm}
A.~F.~Santos, A.~Y.~Petrov, W.~D.~R.~Jesus and J.~R.~Nascimento,
``G\"odel solution in the bumblebee gravity,''
Mod. Phys. Lett. A \textbf{30} (2015) no.02, 1550011
[arXiv:1407.5985 [hep-th]].


\bibitem{Jha:2020pvk}
S.~K.~Jha and A.~Rahaman,
``Bumblebee gravity with a Kerr-Sen-like solution and its Shadow,''
Eur. Phys. J. C \textbf{81} (2021) no.4, 345
[arXiv:2011.14916 [gr-qc]].


\bibitem{Filho:2022yrk}
A.~A.~A.~Filho, J.~R.~Nascimento, A.~Y.~Petrov and P.~J.~Porf\'\i{}rio,
``Vacuum solution within a metric-affine bumblebee gravity,''
Phys. Rev. D \textbf{108} (2023) no.8, 085010
[arXiv:2211.11821 [gr-qc]].


\bibitem{Xu:2022frb}
R.~Xu, D.~Liang and L.~Shao,
``Static spherical vacuum solutions in the bumblebee gravity model,''
Phys. Rev. D \textbf{107} (2023) no.2, 024011
[arXiv:2209.02209 [gr-qc]].


\bibitem{Liu:2024axg}
J.~Z.~Liu, W.~D.~Guo, S.~W.~Wei and Y.~X.~Liu,
``Charged spherically symmetric and slowly rotating charged black hole solutions in bumblebee gravity,''
Eur. Phys. J. C \textbf{85} (2025) no.2, 145
[arXiv:2407.08396 [gr-qc]].


\bibitem{Ovgun:2018ran}
A.~Ovg\"un, K.~Jusufi and I.~Sakalli,
``Gravitational lensing under the effect of Weyl and bumblebee gravities: Applications of Gauss\textendash{}Bonnet theorem,''
Annals Phys. \textbf{399} (2018), 193-203
[arXiv:1805.09431 [gr-qc]].


\bibitem{Oliveira:2021abg}
R.~Oliveira, D.~M.~Dantas and C.~A.~S.~Almeida,
``Quasinormal frequencies for a black hole in a bumblebee gravity,''
EPL \textbf{135} (2021) no.1, 10003
[arXiv:2105.07956 [gr-qc]].


\bibitem{Kanzi:2019gtu}
S.~Kanzi and \.I.~Sakall\i{},
``GUP Modified Hawking Radiation in Bumblebee Gravity,''
Nucl. Phys. B \textbf{946} (2019), 114703
[arXiv:1905.00477 [hep-th]].


\bibitem{Wang:2021irh}
H.~M.~Wang and S.~W.~Wei,
``Shadow cast by Kerr-like black hole in the presence of plasma in Einstein-bumblebee gravity,''
Eur. Phys. J. Plus \textbf{137} (2022) no.5, 571
[arXiv:2106.14602 [gr-qc]].


\bibitem{Islam:2024sph}
S.~U.~Islam, S.~G.~Ghosh and S.~D.~Maharaj,
``Investigating rotating black holes in bumblebee gravity: insights from EHT observations,''
JCAP \textbf{12} (2024), 047
[arXiv:2410.05395 [gr-qc]].


\bibitem{Ali:2024ssf}   
H.~Ali, S.~U.~Islam and S.~G.~Ghosh,
``Shadows and parameter estimation of rotating quantum corrected black holes and constraints from EHT observation of M87* and Sgr A*,''
JHEAp \textbf{47} (2025), 100367
[arXiv:2410.09198 [gr-qc]].


\bibitem{Pantig:2025bpd}
R.~C.~Pantig, S.~Kala, A.~\"Ovg\"un and N.~J.~L.~S.~Lobos,
``Testing black holes with cosmological constant in Einstein-bumblebee gravity through the black hole shadow using EHT data and deflection angle,''
[arXiv:2410.13661 [gr-qc]].

\bibitem{Liu:2019mls}
C.~Liu, C.~Ding and J.~Jing,
``Thin accretion disk around a rotating Kerr-like black hole in Einstein-bumblebee gravity model,''
[arXiv:1910.13259 [gr-qc]].


\bibitem{Jiang:2021whw}
R.~Jiang, R.~H.~Lin and X.~H.~Zhai,
``Superradiant instability of a Kerr-like black hole in Einstein-bumblebee gravity,''
Phys. Rev. D \textbf{104} (2021) no.12, 124004
[arXiv:2108.04702 [gr-qc]].


\bibitem{Wang:2021gtd} 
Z.~Wang, S.~Chen and J.~Jing,
``Constraint on parameters of a rotating black hole in Einstein-bumblebee theory by quasi-periodic oscillations,''
Eur. Phys. J. C \textbf{82} (2022) no.6, 528
[arXiv:2112.02895 [gr-qc]].


\bibitem{Gu:2022grg}
J.~Gu, S.~Riaz, A.~B.~Abdikamalov, D.~Ayzenberg and C.~Bambi,
``Probing bumblebee gravity with black hole X-ray data,''
Eur. Phys. J. C \textbf{82} (2022) no.8, 708
[arXiv:2206.14733 [gr-qc]].


\bibitem{Bluhm:2007bd}
R.~Bluhm, S.~H.~Fung and V.~A.~Kostelecky,
``Spontaneous Lorentz and Diffeomorphism Violation, Massive Modes, and Gravity,''
Phys. Rev. D \textbf{77} (2008), 065020
[arXiv:0712.4119 [hep-th]].

\bibitem{Compere:2008cv}
G.~Compere and S.~Detournay,
``Semi-classical central charge in topologically massive gravity,''
Class. Quant. Grav. \textbf{26} (2009), 012001
[erratum: Class. Quant. Grav. \textbf{26} (2009), 139801]
[arXiv:0808.1911 [hep-th]].


\bibitem{Compere:2009zj}
G.~Compere and S.~Detournay,
``Boundary conditions for spacelike and timelike warped $AdS_{3}$ spaces in topologically massive gravity,''
JHEP \textbf{08} (2009), 092
[arXiv:0906.1243 [hep-th]].


\bibitem{Blagojevic:2009ek}
M.~Blagojevic and B.~Cvetkovic,
``Asymptotic structure of topologically massive gravity in spacelike stretched AdS sector,''
JHEP \textbf{09} (2009), 006
[arXiv:0907.0950 [gr-qc]].


\bibitem{Detournay:2023zni}
S.~Detournay, T.~Smoes and R.~Wutte,
``Boundary conditions for extremal black holes from 2d gravity,''
SciPost Phys. \textbf{16} (2024) no.5, 141
[arXiv:2312.08353 [hep-th]].


\bibitem{Godet:2020xpk}
V.~Godet and C.~Marteau,
``New boundary conditions for AdS$_{2}$,''
JHEP \textbf{12} (2020), 020
[arXiv:2005.08999 [hep-th]].


\bibitem{Cardy:1986ie}
J.~L.~Cardy,
``Operator Content of Two-Dimensional Conformally Invariant Theories,''
Nucl. Phys. B \textbf{270} (1986), 186-204


\bibitem{Anninos:2008fx}
D.~Anninos, W.~Li, M.~Padi, W.~Song and A.~Strominger,
``Warped AdS(3) Black Holes,''
JHEP \textbf{03} (2009), 130
[arXiv:0807.3040 [hep-th]].


\bibitem{Bardeen:1999px}
J.~M.~Bardeen and G.~T.~Horowitz,
``The Extreme Kerr throat geometry: A Vacuum analog of AdS(2) x S**2,''
Phys. Rev. D \textbf{60} (1999), 104030
[arXiv:hep-th/9905099 [hep-th]].


\bibitem{Hofman:2014loa}
D.~M.~Hofman and B.~Rollier,
``Warped Conformal Field Theory as Lower Spin Gravity,''
Nucl. Phys. B \textbf{897} (2015), 1-38
[arXiv:1411.0672 [hep-th]].


\bibitem{Aggarwal:2019iay}
A.~Aggarwal, A.~Castro and S.~Detournay,
``Warped Symmetries of the Kerr Black Hole,''
JHEP \textbf{01} (2020), 016
[arXiv:1909.03137 [hep-th]].


\bibitem{Detournay:2019xgl}
S.~Detournay, W.~Merbis, G.~S.~Ng and R.~Wutte,
``Warped Flatland,''
JHEP \textbf{11} (2020), 061
[arXiv:2001.00020 [hep-th]].


\bibitem{Jiang:2024tmh}
X.~Jiang and J.~Xu,
``Warped CFT duals of the Pleba\'nski-Demia\'nski family of solutions,''
JHEP \textbf{10} (2024), 089
[arXiv:2405.10061 [hep-th]].


\bibitem{Barnich:2007bf}
G.~Barnich and G.~Compere,
``Surface charge algebra in gauge theories and thermodynamic integrability,''
J. Math. Phys. \textbf{49} (2008), 042901
[arXiv:0708.2378 [gr-qc]].


\bibitem{Compere:2007az}
G.~Compere,
``Symmetries and conservation laws in Lagrangian gauge theories with applications to the mechanics of black holes and to gravity in three dimensions,''
[arXiv:0708.3153 [hep-th]].


\bibitem{deBoer:2010ac}
J.~de Boer, M.~M.~Sheikh-Jabbari and J.~Simon,
``Near Horizon Limits of Massless BTZ and Their CFT Duals,''
Class. Quant. Grav. \textbf{28} (2011), 175012
[arXiv:1011.1897 [hep-th]].


	
\end{thebibliography}
}
\end{document}